\documentclass[prx,
amsmath,amssymb,aps,notitlepage]{revtex4-1}

\usepackage{color}
\usepackage{graphicx}
\usepackage{svg}

\usepackage{bm}
\usepackage{bbold}
\usepackage{mathrsfs}
\usepackage{graphicx}
\usepackage{dcolumn}
\usepackage{hyperref}
\usepackage[mathlines]{lineno}
\usepackage{xspace}
\usepackage{textgreek}

\newcommand{\be}{\begin{equation}}
\newcommand{\ee}{\end{equation}}
\newcommand{\bq}{\begin{equation}}
\newcommand{\eq}{\end{equation}}
\newcommand{\ba}{\begin{eqnarray}}
\newcommand{\ea}{\end{eqnarray}}

\newcommand{\lp}{\left(}
\newcommand{\rp}{\right)}
\newcommand{\lb}{\left[}
\newcommand{\rb}{\right]}
\newcommand{\la}{\left\{}
\newcommand{\ra}{\right\}}

\newcommand{\dd}{\mathrm{d}}

\newcommand{\PII}{\mathrm{P}_{\mathrm{II}}}
\newcommand{\PV}{\mathrm{P}_{\mathrm{V}}}
\newcommand{\PVI}{\mathrm{P}_{\mathrm{VI}}}

\DeclareMathOperator{\sech}{sech}
\DeclareMathOperator{\sgn}{sgn}

\newcommand{\ii}{\text{i}}

\newcommand{\bra}[1]{\ensuremath{\langle #1|}\xspace}
\newcommand{\ket}[1]{\xspace \ensuremath{|#1 \rangle}\xspace}

\newcommand{\tilta}{\tilde{\theta}}
\newcommand{\hata}{\hat{\theta}}

\newcommand{\unpm}{\frac{1 \pm m}{2}}

\newcommand{\ps}{\!+\!}
\newcommand{\ms}{\!-\!}

\newcommand{\calD}{{\cal{D}}}

\newcommand{\calH}{{\cal{H}}}

\newcommand{\RR}{{\mathbb{R}}}
\newcommand{\KK}{{\mathbb{K}}}
\newcommand{\ID}{{\mathbb{1}}}

\begin{document}

\title{Universal Painlev\'e VI  
Probability
Distribution 
in   Pfaffian  Persistence and Gaussian First-Passage Problems 
with a sech-Kernel
}
\author{Ivan Dornic}
\email{ivan.dornic@cea.fr}
\affiliation{Service de Physique de l'\'Etat Condens\'e, CEA, CNRS, 
CEA Saclay, 
91191 Gif-sur-Yvette, France}
\affiliation{
Sorbonne Universit\'e, CNRS, 
Laboratoire de Physique Th\'eorique de la Mati\`ere Condens\'ee, 
 4  Place Jussieu,
75005 Paris, France}
\date{\today}

\begin{abstract}
We recast the persistence probability for the spin located at the origin of a
half-space arbitrarily $m$-magnetized Glauber-Ising chain as a Fredholm
Pfaffian gap probability generating function with a sech-kernel. This is then
spelled out as a tau-function for a certain Painlev\'e VI transcendent, the
persistence exponent $\theta(m)/2$ emerging as an asymptotic decay rate. Using
a known yet remarkable correspondence that relates Painlev\'e equations to
Bonnet surfaces, the persistence probability also acquires a geometric meaning
in terms of the mean curvature of the latter, and even a topological one at the
magnetization-symmetric point in terms of Gauss intrinsic curvature.
 Since the same sech-kernel with an underlying
Pfaffian structure shows up in a variety of Gaussian first-passage problems,
our Painlev\'e VI provides their universal first-passage probability
distribution, in a manner exactly analogous to the famous Painlev\'e II
Tracy-Widom laws. The tail behavior in the magnetization-symmetric case of our
full scaling function allows to recover the exact persistence exponent
$\theta(0)/2=3/16$ for the $2d$-diffusing random field or for random real Kac's
polynomials, a particular result found very recently by Poplavskyi and Schehr
[Phys. Rev. Lett. {\bf 121}, 150601 (2018)]. Our Painlev\'e VI tau-function
persistence  distribution also bears a correspondence
with a $c=1$ conformal field theory, the monodromy parameters giving the
dimensions of the associated primary fields. 
This yields $\theta(0)=3 \beta$,
with $\beta=1/8$ the Onsager-Yang magnetization exponent for the critical $2d$
Ising model,  a plausible conjecture relating a nonequilibrium exponent to ordinary static critical
behavior in one more space dimension,  that suggests more generally that methods
of boundary conformal field theory should be helpful in determining the
critical properties of other unsolved nonequilibrium $1d$ processes.
\end{abstract}

\maketitle

\tableofcontents


\section{Introduction and general motivations}

What is the  chance 
for a fluctuating  quantity  to have always remained above its  long-term tendency  or, conversely,  the likelihood 
to first cross its average value
at a given time? 
The study of  the \textit{first-passage} properties
of a random process 
revolves around such questions. It is  a basic problem in probability \cite{Rice, Bendat}, with innumerable applications in the natural sciences \cite{RednerBook,ORReview}.
The usual playground is to consider a centered Gaussian and stationary process $\{Y(T)\}_{T}$, thus uniquely determined by its two-time
translationally invariant correlation function $A(|T_2-T_1|)=\langle Y(T_1) Y(T_2) \rangle$. If  this correlator  vanishes sufficiently fast, physical intuition dictates that $P_0(T)$, the no-crossing probability at zero level ($=\langle Y \rangle$)
and up to a (fixed) time $T$,
should behave, at least for large $T$, 
as $P_0(T) \propto e^{-\theta T}$, with an asymptotic \textit{decay rate} $\theta$.
Leaving aside the Markovian memoryless case, where one can show that
$P_0(T)=\frac{2}{\pi} \arcsin{[A(T)]}$, with  necessarily then $A(T)=e^{-\theta |T|}$ at all times,
 and despite a huge number of works
accumulated over decades, notably in the signal theory or applied probability literature \cite{Slepian62,ZZ69,Shepp71,Li04,Molchan},
 there exists hardly a handful of stationary Gaussian processes for which 
the decay rate $\theta$  has been determined without approximations,
and still less so for the full first-passage probability  $-\dd P_0(T)/\dd T$. Such quantities, that embody  infinite-order correlation functions,
 are sensitive to the whole history of the process, as encoded into finer analytical details of the correlator. Without an additional structure present or some approximations made, there is therefore  little hope to be able to calculate them.

Starting from the nineties, studies of simple models of phase-ordering kinetics have triggered a renewed interest on first-passage properties
under the wording of \textit{persistence} \cite{SNM99,BMS13,AS15}. In the 
context of many-body interacting
nonequilibrium systems, such as those displaying coarsening \cite{Bray94},
 the issue is to understand how  a local degree of freedom
can maintain its initial orientation
as  domains of globally aligned spins form and grow forever.
Taking for simplicity $\pm$ Ising spins  quenched from a disordered initial state to 
zero temperature, the persistence probability $p_0(t)$ is then defined
as the probability
that a given spin has always remained in its initial state up to time $t$. This also has  a
\textit{geometric} meaning, being the fraction of  spins which have never flipped up to time $t$, while $-\dd p_0(t)/\dd t$ is the probability
that a domain wall first sweeps over a particular location in space.
Part of the initial activity on this subject, and the fascination for it,
was triggered by the discovery that even for the simplest possible
models \cite{DBG94,DHP95,MSBC96,DHZ96,DHP96},
the algebraic decay $p_0(t) \propto t^{-\theta}$ takes place  
with an
exponent $\theta$ which does not seem to be related to  other known static or dynamic critical exponents, yet subsuming in a simple number
the everlasting 
evolution of the  interwoven mosaic of coarsening domains.

Early studies 
culminated in two apparently unrelated climaxes. 
On  one hand it was the finding \cite{MSBC96,DHZ96} that  for the simple diffusion equation evolving from random initial conditions  --- a popular model of phase-ordering 
where one considers the sign of the (Gaussian) diffusing field as the local spin variable \cite{OJK82} --- the persistence exponent is a non-trivial function $\tilde{\theta}(d)$ in any space dimension $d$, 
unrelated to the generic diffusive lengthscale $L(t) \propto t^{1/2}$. With hindsight, this is less surprising \cite{SciencePop}, since the  algebraic
decay $p^{{\mathrm{dDiff}}}_0(t) \propto t^{-\tilde{\theta}(d)}$  of the persistence probability  merely
corresponds, when viewed  on a
logarithmic timescale $T=\ln{t}$ where the statistical self-similarity of coarsening domains in the scaling regime appears  stationary, to the decay rate 
  for the no-crossing probability $P^{{\mathrm{dDiff}}}_0(T)= p^{{\mathrm{dDiff}}}_0(e^T)\propto e^{-\tilde{\theta}(d) T}$ of a stationary Gaussian process with a never Markovian correlator  \cite{footnote_DiffMarkov}.
 Later, it has also been realized that the  particular case of the 
  $d=2$ diffusion equation plays a distinguished role. 
  Indeed, if one considers 
  the  Gaussian instantaneous value $X(t)=\varphi(0,t)$ of the local diffusing field
  $\partial_t \varphi=\nabla^2_{r} \varphi$ (at $r=0$  without restriction  
  by  translational invariance of the Gaussian random initial condition),  and the associated normalized 
  process $Y(T)=X(e^T)/[\langle X^2(e^T)\rangle]^{1/2}$ (still Gaussian), then
  the corresponding 
  correlator,
  $A(T)=\langle Y(0) Y(T) \rangle=1/\cosh{(T/2)}$, a smooth, even function of $T$,
 appears in one guise or in another into questions as diverse as the determination of the probability that a random Kac polynomial has no real roots, or that of the gap probability for eigenvalues of truncated random orthogonal matrices \cite{DPSZ02,Li04,SM07,SM08,PJF10,MS13, DM15}. For all these Gaussian problems that 
 share the same correlator and  are thus actually identical (at least within some
 appropriate scaling regime),
 the persistence exponent  $\tilde{\theta}(2)=0.1875(10)$ (taking the "best" numerically-determined value from \cite{EMB04}) therefore represents 
 the universal 
 decay rate 
 of their common  first-passage probability.
 
  On the other hand, and in a  genuine \textit{tour-de-force},
the authors of \cite{DHP96}
managed to provide an exact expression (valid at all times)
for  the more general persistence 
probability $p^{{\mathrm{Potts}}}_0(t_1,t_2;q)$ 
that a  "colored" $q$-state Potts spin on a $1d$ chain evolving under zero-temperature Glauber dynamics has never flipped between times $t_1$ and $t_2$, and to
extract (for large $t_2/t_1$) the 
 corresponding algebraic decay $p^{{\mathrm{Potts}}}_0(t_1,t_2;q) \propto (t_2/t_1)^{-\hata(q)}$. 
 Pivotal in the derivation of their results is the
 consideration of the
 persistence probability $p_0^{\mathrm{SemiP}}(t_1,t_2;q)$ in a specific geometry, the semi-infinite  chain, and for the particular spin located at the origin here. This suffices to reconstruct the full persistence probability
  on an infinite chain, 
    because  if any  spin there does not flip,
the motion of the domain walls to the left and to its right occurs independently. Conversely, 
$p_0^{\mathrm{SemiP}}(t_1,t_2;q)= [p_0^{\mathrm{Potts}}(t_1,t_2;q)]^{1/2}$,  thus  decaying with a halved exponent.
The crucial observation of \cite{DHP96} is that $p_0^{\mathrm{SemiP}}(t_1,t_2;q)$
 can be obtained by uncovering a certain algebraic structure, a \textit{Pfaffian}. This allows --- and this only for the 
 particular spin  located
at the origin of the semi-infinite chain --- to express the 
probability that it has 
the same value at an arbitrary number of fixed times just in terms 
of combinations of a basic building block, constituted by the
two-body, no-meeting probability between a pair of random walkers.
   Once the dust  has settled,  after many more technical hurdles to be overcome,
   the eventual upshot is a  persistence exponent $\hat{\theta}(q)$ with a complicated but explicit expression \cite{DHP95,DHP96},
 never rational
for any finite  $1 < q < \infty$, except for $q=2$ Ising spins,
for which  one finds 
the strikingly simple number 
$\hat{\theta}(2)=3/8$.
Yet one cannot 
help but muse over the puzzling 
numerical proximity between the thereby determined persistence exponent for the semi-infinite Ising chain, 
$\hat{\theta}(2)/2=0.1875$,  and the value 
$\tilde{\theta}(2)=0.1875(10)$ found
for the
$2d$-diffusing random field. But how these two model systems, apparently
so dissimilar, and that do not even live in an ambient space with the same 
physical dimension, could possibly be related at the level of a
quantity so sensitive to details as the persistence exponent?

\section{Summary of  results}

In this note we demonstrate, among other things, how one can answer such a question.
Our central result is that, when viewed in the scaling regime on the logarithmic timescale $T$ where the coarsening is stationary,  the  
persistence probability $P_0^{\mathrm{HalfI}}(T;m)$ 
for the particular $\pm$ Ising spin located
at the origin of an arbitrarily $m$-magnetized half-space chain--- a quantity simply related to $p_0^{\mathrm{SemiP}}(t_1,t_2;q)$ for
$q$-state Potts spins and thus sufficient
to rebuild up the full persistence
probability --- 
 can be recast
at all times $T$ 
as a {\it universal}
probability distribution involving a tau-function for a member of the highest hierarchy of
Painlev\'e transcendents, a Painlev\'e VI ($\PVI$). The particular $\PVI$
which shows up is the function $y=y(x)$ \textit{defined}  by the solution of 
\be
\label{P6}
 \frac{\dd^2 y}{\dd x^2} = \frac{1}{2}\lp \frac{1}{y} \ps\frac{1}{y-1} \ps \frac{1}{y-x} \rp \lp \frac{\dd y}{\dd x}\rp^2
\ms \lp \frac{1}{y} \ps \frac{1}{y-1} \ps \frac{1}{y-x} \rp \lp \frac{\dd y}{\dd x}\rp  \ps 
\frac{y(y - 1)(y - x)}{x^2(x - 1)^2} \lp 
\frac{1}{8}  \frac{x-1}{(y-1)^2}+\frac{3}{8}
 \frac{x(x-1)}{(y-x)^2}
\rp,
\ee
with striking numerical values for its parameters, visible in  and determined from the non-differential part.
The independent variable $x$ of 
Eq.~(\ref{P6}) in rational coordinates corresponds to  $x=(t_2/t_1)^2=e^{2 T}$, 
 with $T=\ln{(t_2/t_1)}$ the logarithmic timescale for the persistence problem. 
 Therefore just the branch  $y(x)$  of the solution to  Eq.~(\ref{P6}) with $x \in (1,+\infty)$ is relevant.
The monodromy parameters  of our $\PVI$, the four $\{\vartheta_\nu\}_{\nu}$, (a customary notation
for $\PVI$, yet an unfortunate one in the context of  persistence),
 that encode (up to a sign)
 the eigenvalues $\pm \vartheta_{\nu}/2$ around each of the
 four fixed singularities
 $\nu=0,1,x,\infty$ (and up to any homographic transformation permuting their locations),
  are therefore:
\be
\label{monom}
\la \vartheta_\infty^2, \vartheta_0^2, \vartheta_1^2,\vartheta^2_x \ra = \la 0,0,\frac{1}{4}. \frac{1}{4} \ra,
\ee
 Notice also that even though there exists a choice of signs that makes the sum $\sum_{\nu}\vartheta_\nu$ an integer,
 corresponding to the classical hypergeometric solutions for the $\PVI$
 equation \cite{Handbook},
 one can check that the required  physical solution to define a properly normalized
persistence  probability distribution is a genuinely transcendental $\PVI$.

We now briefly sketch the main steps of our derivation, and their consequences. It uses extensively   methods developed in the context of Random Matrix Theory (RMT), although it does not
take advantage of the structure imposed {\it a priori} by some random matrix ensembles. It turns out however
that lurking in the background is the Ginibre Orthogonal Ensemble (GinOE) for real random matrices,
usually considered as the most complicated of all standard RMT ensembles \cite{MehtaBook,Ginibre}, and that has already appeared
in the exact determination of the distribution of domain sizes for the $1d$-Potts model \cite{DZ96,FGinOE2015}. The precise relationship between the complex-valued Pfaffian point process formed by the eigenvalue distribution of the GinOE  and the computation of temporal properties such as the persistence 
is another (long) story, that we figured out only when this work was already under completion. This will be discussed elsewhere \cite{IDPrep1}. Notice also that the sech-kernel, Eq.~(\ref{defK}), at the heart of this work, can be  viewed as a very particular case of a much more general hypergeometric kernel, that had appeared in the
representation theory of some big,  infinite-dimensional groups \cite{Big02},
figured out there through Riemann-Hilbert techniques. We do not use any such high-level ideas or tools here, and our approach will be a much more bottom-up, pedestrian one.
At the end of our journey, beyond sizable technical and conceptual
difficulties,  the main take-home message
is simple. It is that on the exemplary value of the  persistence probability, Painlev\'e transcendents, that
reside at the crossroad of so many branches of
mathematics, are capable of fulfilling the Holy Grail of statistical physics: the exact integration
through a  non-Markovian temporal process
of the spatial degrees of freedom in a many-body
 interacting system. 

\section{Persistence as a Fredholm Pfaffian gap probability generation function}

The first part of our analysis is to identify that, in the most convenient variables $(T=\ln{(t_2/t_1)},m=-1+2/q)$, 
    the expression
    obtained in \cite{DHP96} for the persistence probability $p_0^{\mathrm{SemiP}}(t_1,t_2;q)$,  can be recast as a Fredholm Pfaffian gap probability generating function (with parameter $\xi=1-m^2$) for 
    the restriction $K_T$ to the symmetric interval $[-T/2,T/2]$ of a certain  continuous integral operator $K(x,y)$ dubbed, for obvious reasons, the \textit{sech-kernel}:
 \bq
 \label{defK}
 K(x,y)= K(x-y)= \frac{1}{2 \pi} \sech{[(x-y)/2]},
 \ee
 (recall that $\sech=1/\cosh$).
This kernel corresponds precisely 
 to the stationary Gaussian correlator $A(T)=\rho_0^{-1} K(T)$ for the $2d$-diffusing random field,   $\rho_0=1/(2 \pi)$ being  the average density of zero-crossings
 for the latter, i.e. the inverse of the mean spacing distance (on the logarithmic timescale) 
 between changes of sign of $\{Y(.)\}$  \cite{MSBC96,DHZ96}.
 The crucial quantities to consider to show this are $P_0^{\pm}(T;m)$,  the   probabilities for the Ising spin located at the origin of a half-space $m$-magnetized chain to have always been in the $+$ state (resp. $-$) during a length of logarithmic time $T$. These two persistence probabilities obey  simple symmetry and sum-rule relations, both for all $T \ge 0$ and $m \in (-1,1)$:
 \begin{eqnarray}
 \label{sym}
 P_0^-(T;m) &=&P^+_0(T;-m), \\
 \label{sum}
 P_0^+(T;m)&+&P^-_0(T;m) =P_0^{\mathrm{HalfI}}(T;m).
 \end{eqnarray}
 Eq.~(\ref{sym}) is obtained by reversing globally the initial condition, and that does not affect the dual dynamics of the coalescing random walkers determining the probability that the spin
 at the origin has not flipped, while Eq.~(\ref{sum})
 is the full persistence probability for the Ising spin located at the origin of the half-space $m$-magnetized chain, irrespective of the state in which it stayed for a length of time $T$.
 Thus one of  these two probabilities, say $P^+_0(T;m)$  determines everything, and it is
 then related  \cite{footnote_Subtle} to the central result  obtained in \cite{DHP96} (Eq.~(29) there) for the persistence
 probability $p_0^{\mathrm{SemiP}}(t_1,t_2;q)$ of the $q$-state Potts spin at the origin of a semi-infinite chain by:
 \be
 \label{relPp}
 P^+_0(T=T_2-T_1;m)=\frac{1}{q}p_0^{\mathrm{SemiP}}(e^{T_1},e^{T_2};q)\Big{|}_{1/q=(1 \ps m)/2}.
 \ee
 The factor $1/q$ on the right-hand-side  of Eq.~(\ref{relPp}), with $1/q=(1+m)/2$, comes from
 relabelling a particular Potts color, occurring with probability $1/q$ in the random initial condition, as a $+$ Ising spin, while the $q-1$ other colors are lumped together as $-$ Ising spins, induced therefore an overall
 magnetization $m=1/q-(1-1/q)$ for this semi-infinite or \textit{half-space} $m$-magnetized Ising chain. The equivalence of these two viewpoints with a random initial condition in both cases is well-known, see e.g \cite{DZ96}.
 The way the persistence exponent $\hata(q)$ was then extracted in \cite{DHP96} from the right-hand-side of Eq.~(\ref{relPp}) is from 
 Szeg\"o-Kac-Akhiezer formula giving
 the asymptotic behavior of truncated Wiener-Hopf operators (the continuous version of Toeplitz determinants). The   continuous kernel which shows up originates from taking the limit of a dense set of times
  for the spin not to flip between times $t_1$ and $t_2$, and  can be derived from
  the basic Pfaffian building block formed
  by the probability $c(s,t)$ that
 pair of $1d$ random walkers starting from the origin at times $s < t$ and
 going backwards in time while wandering on the positive half-space
 have not met down to the initial condition. 
 Using the scaling form $c(s,t) \approx (2/\pi)(\arctan{\sqrt{t/s}}-\arctan{\sqrt{s/t}})$  (Eq.~(15) of \cite{DHP96}, rewritten), we observe now that the kernel appearing in \cite{DHP96} is
 simply related to the sech-kernel in the logarithmic variables $s=e^{x},t=e^{y}$, with $1/q=(1\ps m)/2$:
 \be
 \label{relMK}
 \lb \delta(s \ms t) \!+ \!2\frac{q-1}{q^2} \frac{\dd}{\dd s}c(s,t)\rb \dd t= \lb \delta(x\!-\!y)\!-\!(1\!-\!m^2)K(x\!-\!y) \rb \dd y.
 \ee
 Equipped with Eq.~(\ref{relPp}) and Eq.~(\ref{relMK}), our original contribution then amounts to  rewrite in the $(T,m)$-language, and  following the framework of \cite{TWGOSE} valid for any even difference kernel on a symmetric interval, the (complicated) expression found in \cite{DHP96} for the amplitude of the right-hand-side of Eq.~(\ref{relPp}). 
 The outcome is that
 the two persistence probabilities $P_0^{\pm}(T;m)$ can be expressed in a remarkably compact manner
 in terms of the even and odd parts, $\calD_{e,o}(T;\xi)$, of the Fredholm determinant  $\calD(T;\xi)=\prod_{k=0}^{\infty}[1-\xi \lambda_k(T)]$ for $\xi K_T$, with $\xi=1 \ms m^2$, viz. 
 \bq
\label{resP0plus}
P_0^{\pm}(T;m) = \frac{\calD_e(T;1 \ms m^2) \pm m \, \calD_o(T;1\ms m^2)}{2},
\eq
an expression holding true down to $T =\ln{(t_2/t_1)} \to 0^+$, where it identifies with $\unpm$,  the probability
for a spin to be either $\pm$ on a  $m$-magnetized chain (any arbitrary initial magnetization being preserved by zero-temperature Glauber dynamics at all subsequent times \cite{Glauber63}). 
An immediate consequence of Eq.~(\ref{resP0plus}), due to the  sum-rule Eq.~(\ref{sum}) for the two persistence probabilities, is that
\be
\label{resP0I}
P_0^{\mathrm{HalfI}}(T;m)=\calD_e(T;1 \ms m^2) =
\exp{\la \! - \!\int_0^{T/2}\! \! \! \dd x [R(x,x)\! + \! R(x,-x)] \ra},
\eq
where we have used for the last equality a well-known formula (see e.g. \cite{TWIntro}), giving a Fredholm determinant in terms of the matrix elements of the \textit{resolvent} operator $\RR$ for $\xi K_T$ (with still $\xi=1 \ms m^2$). We recall that  the resolvent as an operator is defined by $\ID+\RR=(\ID-\xi K_T)^{-1}$, 
and we denote its  matrix elements 
for $x,y \in [-T/2,T/2]$ as
 $R(x,y)=R(x,y;\xi)=\bra{x} \RR \ket{y}$ (using Dirac's bra-ket notations, with in particular $\delta(x-y)=\bra{x}\ID \ket{y}$).
Eq.~(\ref{resP0I}) is therefore the Fredholm
determinant generating function for the
 even part $K_e$ of the sech-kernel: $K_e(x,y)=(K(x-y)+K(x+y))/2$.
 The whole derivation  can actually be rendered independent of the precise form Eq.~(\ref{defK}) of the kernel, a fact already hinted at in \cite{DHP96} when leaving the expression of $\hata(q)$ in terms of the scaling form of $c(s,t)$ in the original time variables. It just depends on the intrinsic Pfaffian structure of the persistence probability, and  an equation such as Eq.~(\ref{resP0I}) holds true when expressed
 in the logarithmic variables for   any even-difference kernel on a symmetric interval \cite{MehtaBook,TWGOSE}. In fact,   Eq.~(\ref{resP0I}) is the pristine analog for the sech-kernel of a  celebrated result obtained 
 in the context RMT   by  Gaudin for the  sine-kernel  \cite{Gaudin61}, who   computed in terms of the (even part) of the corresponding
 Fredholm determinant
 the exact spacing distribution
 in the "bulk" scaling limit of the Gaussian Orthogonal Ensemble  (GOE), showing thereby that
 Wigner's surmise was indeed a very good approximation for real random matrices.
 Twenty years after, it was recognized by the Kyoto School, although in a somewhat terse manner \cite{MehtaBook,Mehta92} at the detour of a long and profound article on monodromy preserving deformations  \cite{Bose80}, that this very same gap spacing 
 Fredholm determinant can be expressed as a  \textit{tau-function}, involving the non-autonomous Hamiltonian evaluated on the equations of
 motion
 associated to a certain Painlev\'e V ($\PV$) function.
 
 \section{Computation of Fredholm determinants for the integrable sech-kernel}
 
    The second main step of our analysis is to perform the genuine computation of the Fredholm
    determinants for the sech-kernel in terms of which the persistence probability is expressed.
    A difference with the standard kernels of RMT is that there does not seem to exist here a bilinearization formula, that  permits to find a differential operator $L$ commuting with  the sech-kernel,
    and thus to compute the Fredholm determinant for $K_T$ with the help of the eigenfunctions of $L$, as was done with  prolate spheroidal functions for the bulk GOE  in \cite{Gaudin61}, or for the Airy and Bessel kernels \cite{TWAiry,TWBessel}.
    Yet if one rewrites the 
 sech-kernel with the help of the innocent-looking, algebraic identities: 
\bq
\label{Integrable}
\frac{1}{\cosh{[(x \!- \! y)/2]}}= \frac{2 \sinh{[(x \! - \!y)/2]}}{\sinh{[(x \!- \!y)]}}= \frac{e^{3 x/2} e^{y/2}-e^{x/2}e^{3 y/2}}{e^{2 x}-e^{2 y}},
\eq
one  recognizes that it gives rise to a so-called integral \textit{integrable} operator \cite{IIKS90}, to which one can still apply
the formidable functional-analytic machinery developped by Tracy and Widom in the nineties to compute the corresponding Fredholm
determinant for $K_T$ as a function of the endpoints of the interval where it is defined \cite{TWAiry, TWBessel, TWFred, TWGOSE, TWIntro}.  The situation here even corresponds to a case briefly considered in \cite{TWFred}
for  the 
finite-matrix size $N$ sine-kernel of the Circular Unitary Ensemble (CUE) of RMT, $K_N(x,y)=(1/N)\sin{[N(x\!-\!y)/2]}/\sin{[(x\!-\!y)/2]}$, up to some analytic continuation $x,y \to 2 \ii x, 2 \ii y$, and yet with $N=1/2$! Incidentally, this very observation provided the impetus some years ago for this work, since it is known from the existing literature \cite{WFC2000,FW2002,FW2004} that the finite-$N$ CUE gives rise to a $\PVI$ transcendent where
the matrix size just appears as a parameter in its monodromy exponents, and which coalescences when $N \to \infty$ to the Jimbo-Miwa-M\^ori-Sato $\PV$  associated to
the sine-kernel. We thereby had a clue for the possible occurrence of a  non-trivial Painlev\'e transcendent associated to the sech-kernel,
hidden somewhere between the above two in the hierarchy, up to some analytic continuation that ought to be relatively innocuous, since these functions are mathematical objects living
intrinsically in the
complex (projective) plane.
At any rate, using the exponential Christoffel-Darboux decomposition provided by Eq.~(\ref{Integrable}), one can follow the nearly algorithmic and by now well-trodden footsteps of \cite{TWFred}.
After differential elimination of intermediate quantities, one arrives at a \textit{closed}
 second-order second-degree nonlinear ODE for $H(T)=R(T/2,T/2)$, the resolvent function at coincident
points, that we  express in terms of the physical variable $T$, i.e. the length of the interval where our kernel acts:
\be
\label{ODEH1}
\lp \frac{H''}{H'}+2 \coth{T} \rp^2 -4\lp \frac{H^2}{H' \sinh^2{T}} + 2 H H' \coth{T} + H' \rp =1.
\eq
[Compare Eq.~(\ref{ODEH1}) with Eq.~(5.70) of \cite{TWFred}: they do indeed match with $t \to \imath T/2$  and $N^2=1/4$, as anticipated.]
As in studies of standard RMT kernels, the only place where appears the $\xi$-dependence of the resolvent for the diluted kernel $\xi K_T$
is in the initial condition: by the Neumann expansion near $T=0$ of the resolvent, one finds that $H(0)=\rho_0(1 \ms m^2)$ (that can already  be seen at the level of Eq.~(\ref{relMK})). Thereby, from Eq.~(\ref{ODEH1}), order by order
  all the coefficients of the Taylor-expansion for $H(T)$ near the origin can be expressed in terms of 
$H(0)$, with in particular $H'(0)=H^2(0)$ (cancelling the spurious pole at the origin due to our rewriting of Eq.~(\ref{ODEH1})).
Therefore there should exist a unique  regular solution to Eq.~(\ref{ODEH1}) on the whole positive real axis with
local Cauchy initial datum $\{H(0)=\rho_0(1 \ms m^2), H'(0)=H^2(0)\}$, joining one  having a  \textit{finite} limiting value as $T \gg 1$, this last condition being required through Eq.~(\ref{resP0I}) by the very existence of a persistence
exponent. In other words, the formula for  the persistence
exponent should be rigidly determined by the analytic structure.
 By rewriting  in terms of $m$ and in a somewhat more 
symmetric form the expression found 
in \cite{DHP95,DHP96} for $\hata(q)$, we recast the corresponding expression for $\theta(m)$ as:
\bq
 \label{thetam}
 \theta(m)= \hata \lp q=2/(1\ps m)\rp=
 \frac{1}{2}\la 
 \lb \arccos{\lp \frac{m}{\sqrt{2}}\rp} \rb^2
 - \lb \arccos{\lp \frac{1}{\sqrt{2}}\rp} \rb^2
 \ra.
 \ee
 Notice that $\theta(m)$ is \textit{not} an even function of $m$.
This regular solution $H(T)=H(T;H(0),H^2(0))$  also permits to define (for a fixed value of $m$) two well-normalized  probability density functions $p_1,p_2$ on $(0,+\infty)$,  with (tail) probability distribution functions  $F_1,F_2$. The first of these two 
is simply the Fredholm determinant for $(1-m^2) K_T$:
\be
\label{F2}
 F_2(T) = \det{[\ID \!-\!(1\!-\!m^2) K_T]}= \int_{T}^{\infty} \! \! \dd \ell \, p_2(\ell)= \exp{\lb-\int_0^T \! \dd \ell \, H(\ell)\rb},
\ee
while the second is a rewriting of Eq.~(\ref{resP0I})
for the persistence probability, using the so-called Gaudin's relation $\dd R(x,x)/\dd x=R^2(x,-x)$,  again valid for any even difference kernel \cite{MehtaBook, Mehta92, TWGOSE}:
\be
\label{F1}
 F_1(T) = P_0^{\mathrm{HalfI}}(T;m)= \int_{T}^{\infty} \! \! \dd \ell \, p_1(\ell)=
 \lb F_2(T) \rb^{1/2} \exp{\lb-\frac{1}{2}\int_0^T \! \dd \ell  \sqrt{H'(\ell)} \rb}
\ee
Thus the knowledge of $F_2$, i.e. that of $H$ (and of its derivative) is sufficient to determine
$F_1$. Notice that, obviously, $F_2(0)=1$, and $F_2(T) \propto e^{-\theta(m)T} \to 0$ as $T \to \infty$, since $H(T) \to \theta(m)$ there. Plots of the distribution $F_2$ and of its density should probably more accessible from the Fredholm determinant
representation Eq.~(\ref{F2}), rather than solving directly the ODE Eq.~(\ref{ODEH1}) \cite{Bornemann2010}.
As for the distribution $F_1$, namely the persistence probability for a fixed value of $m$, 
its behavior for large $T$ is more subtle, since
\be
\label{F1decay}
F_1(T) \propto
\exp{\la-\mathrm{min}[\theta(m),\theta(-m)]T/2 \ra},
\ee
because  either $P_0^+(T;m) \propto e^{-\theta(m)T/2}$ or $P_0^-(T;m) \propto e^{-\theta(-m)T/2}$ dominates the sum Eq.~(\ref{sum})
according to the sign of $m$ (the relation for the respective decay rates coming from Eq.~(\ref{sym})), with balanced
contributions only for symmetric initial conditions. By varying $m$ one therefore 
observes a cusp at the
origin for the overall decay rate of $F_1(T)$. Yet one can explore the full range of values for
$\theta(m)$ given by Eq.~(\ref{thetam}) by following continuously one of the two branches $\theta(\pm m)$, and 
that was precisely what has been done for the corresponding $\hata(q)$ in \cite{DHP96}, or simply by a further conditioning of $P_0^{\pm}(T;m)$ with respect to the value the spin had at $T=0$.
This being said, for a fixed value of $m$,
 Eqs.~(\ref{F2})-(\ref{F1})
 are the  analogs
for our  sech-kernel to the  famous 
Painlev\'e II ($\PII$) Tracy-Widom 
distributions \cite{TWAiry}, defined
in terms of Fredholm determinants for the Airy kernel,
and that give respectively the (inter-related) distributions of the maximum eigenvalue in the $\beta=2$  GUE and the 
 $\beta=1$ GOE when properly scaled at the soft edge at the spectrum. Eq.~(\ref{ODEH1}) is  the equivalent (for each value of $m$) to  the $\PII$ Hastings-McLeod solution 
that appears in the Tracy-Widom distributions.

\section{The Painlev\'e VI Bonnet surface}

{\textAlpha \textgamma \textepsilon \textomega \textmu \textepsilon \texttau \texteta \texttau  \textomikron \textzeta \quad \textmu \texteta \textdelta \textepsilon \textiota \textzeta \quad \textepsilon \textiota \textsigma \textiota \textiota \texttau \textomega}. (\textit{Let no one who cannot think geometrically enter here}.) 


[Legendary inscription written at the entrance of Aristotle's classroom in Plato's Academy.]
\vspace{0.25cm}

The third stage of our work consists in the
precise identification of the underlying $\PVI$
function $y(x)$ for which the above probability
distribution 
$F_2$ is a  tau-function,
and especially the determination of
its four associated monodromy exponents.
Given the 2nd-order 2nd-degree nonlinear ODE satisfied by $H(T)$, there are many ways to
do this correspondence \cite{CG93,Handbook}, and to express the rational 
transformation relating a $\PVI$ function $y(x)$
and its derivative $y'(x)$
to one of its
Hamiltonian $H(T) \propto \calH_{\mathrm VI}(q(s),p(s),s)$ evaluated on the equations of motion, with appropriate relations between the respective independent variables $x,s,T$ in each case.  These are straigthforward methods, although a bit heavy \cite{CG93}. Here we shall use a marvelous geometric shortcut --- a remark that we owe to Robert Conte \cite{RCPriv} --- that permits to recognize in   our equation established for $H(T)$, Eq.~(\ref{ODEH1}), the so-called Hazzidakis equation, that appeared
as early in 1897 \cite{Hazzidakis},
in a question of  differential geometry of surfaces
stated (and answered) by Pierre-Ossian Bonnet in 1867 \cite{Bonnet1867}!
Given the two fundamental Gauss' quadratic forms of a surface defined in conformal coordinates ($z,\overline{z})$, how to determine
all the surfaces that are applicable on each other? It turns out that, as shown by Bonnet, their \textit{mean curvature} function $\mathscr{H}_M$ has to satisfy a certain third-order nonlinear ODE depending just on $\Re{z}$, for which Hazzidakis found
an integral of motion, of which the right-hand-side of Eq.~(\ref{ODEH1}) is
a particular case. It took then  one more century until
 Bobenko and Eitner \cite{BE94} identified 
 Hazzidakis' equation for  $\mathscr{H}_M(\Re{z})$ as
a  Hamiltonian for a certain one-parameter $\PVI$. The latter remarkable correspondence uses the fact that the Gauss-Codazzi
equations for the moving frame of a Bonnet surface can be retranscribed
as a Lax pair for $\PVI$ \cite{BEK95,BEbook}.
As shown in 2017 by Robert Conte \cite{RC16,RC17}, 
the corresponding codimension-three Bonnet-$\PVI$ can be extrapolated to the "full" $\PVI$ with
four arbitrary monodromy parameters to obtain, due to its geometric origin, probably the "best" (more symmetric) Lax pair. 

For our particular case, this  rather unexpected incarnation of the coarsening motto, according to which the motion of the domain walls proceeds by mean curvature, yields the explicit
relation $H(T)=-2\mathscr{H}_M(\Re{z})$ (compare our Eq.~(\ref{ODEH1}) with Eq.~(B.20) from \cite{RC17}), with $\Re{z}=T$, and the specific Bonnet-$\PVI$, Eq.~(\ref{P6})
for $y=y(x)$, with $x=e^{2 T}$, and its accompanying
monodromy exponents, Eq.~(\ref{monom}) through:
\be
\label{Hy}
H(T)= -\frac{(x - 1) x^2}{y(y - 1)(y - x)}
\lp \frac{1}{2} \lp\frac{\dd y}{\dd x}\rp^2 - \frac{1}{8}\frac{y^2}{x^2(y - x)}\rp.
\ee
Hence $F_2(T)$ is indeed a $\PVI$-tau function.
We mention that this implies that a formula equivalent to Eq.~(\ref{thetam})  is actually buried as early in 1982 in
the so-called Jimbo-Fricke cubic of monodromy invariants exhibited in \cite{Jimbo82}, since there  the full  connexion problem
for the tau-function of $\PVI$  was solved
 \cite{RCIDPrep}. This also gives a precise meaning to both  intuitive viewpoints advocated in the introduction of the persistence exponent as a
decay rate endowed with a geometric interpretation, the former
aspect   for an admittedly complicated
time-dependent $\PVI$ Hamiltonian 
 evaluated on the equations of motion, yielding nevertheless
 a sort of
(exact) Kramers' formula.  A  tau-function characterization  holds also true for $F_1(T)$, with additional significant complications \cite{RCPriv,RCIDPrep}.
 
 This geometric correspondence also gives  a flurry of results, among which
a precise relationship between the two asymptotic  curvatures $\kappa_1,\kappa_2$ of the underlying Painlev\'e-Bonnet (saddle-like) surface  and the persistence exponent  as $T=\Re{z} \to \infty$:
\be
\label{k1k2tm}
\frac{\kappa_1 + \kappa_2}{2}= -\frac{\theta(m)}{2}.
\ee
Eq.~(\ref{k1k2tm}) thus provides an explanation for the appearance of angles in Eq.~(\ref{thetam}), as
a sort of non-linear quadratic Buffon's needle formula (see \cite{EK95} for a beautiful exposition of this problem in the spirit of modern random geometry, as well as, incidentally,  for a relation with the zeros of Kac's polynomials, as testified by the very title of that work).
Particularly noticeable are the consequences
of this Painlev\'e-Bonnet correspondence
 for the distinguished value  $m=0$
 of the persistence exponent. For  Bonnet surfaces, it turns out that
 the mean curvature function also determines its first fundamental quadratic
 form, a.k.a. the metric, through:
 \be
 \label{metric}
(\dd s)^2 = \frac{\dd z \dd \overline{z}}{H'(T) \sinh^2{T}},
\ee
with still $T=\Re{z}$, while $\Im(z)$ is related to the spectral parameter  in the Lax pair representation of the moving frame \cite{BE94,RC17}, the piece $1/\sinh^2{T}$ in Eq.~(\ref{metric} being the so-called Hopf factor of the surface. From Eq.~(\ref{metric}), the metric
therefore becomes singular  when $H'(T)$ vanishes. This corresponds
to a so-called \textit{umbilic} point, where the two local curvatures coincide, and where therefore the mean curvature
$\mathscr{H}_M=(\kappa_1+\kappa_2)/2$ is also identical to the square-root of Gauss' curvature, 
$\mathscr{K}_G=\kappa_1 \kappa_2$. The latter is an intrinsic quantity for a surface, independent from any local conformal reparametrization, a result that Gauss called his \textit{Theorema Egregium}
("Remarkable Theorem"). 
A vanishing of the derivative of the mean curvature function is precisely what happens for $m \to 0$, because for large $T$:
\be
\label{singamp}
\exp{\lb \! - \!\int_0^{T/2}\! \! \! \dd x \, R(x,-x) \rb}=\exp{\lb-\frac{1}{2}\int_0^{T}\! \dd \ell  \sqrt{H'(\ell)}\rb} \to \sqrt{|m|},
\ee
 this after carefully  transcribing in the $(T,m)$-language the singular behavior
$\propto \sqrt{q(2-q)}$ for $q \to 2^-$ found in \cite{DHP96} for the asymptotic amplitude of  $p_0^{\mathrm{SemiP}}(t_1,t_2;q)$. This was one of the significant technical hurdle that had to be overcome there, in particular to show by a careful analytic continuation that $\hata(q)$ maintained the same expression for $q>2$, as already hinted at. The above loss of differentiability around $m=0$ can actually be traced back to
the fact that the largest eigenvalue of the Fredholm integral equation for the sech-kernel $\lambda_0(T) \to 1$ as $T \gg 1$, a value corresponding to the maximum  of its (self-dual) Fourier transform 
$\int_{\mathbb{R}}\! \dd v \, e^{-\imath k v} K(v)=\sech{(\pi k)}$, itself attained at $k=0$.
The last phenomenon is also responsible for the "cuspy" behavior evidenced by Eq.~(\ref{F1decay}).
The reward for these two facets of  singular behavior is 
a gem constituted by an intrinsic, topological meaning \textit{\`a la Gauss-Bonnet} for the persistence exponent of  Ising spins with symmetric initial conditions:
\be
\label{kappa}
 \mathscr{H}_M = \sqrt{\mathscr{K}_G}= \kappa=-\frac{\theta(0)}{2}=-\frac{3}{16}.
\ee

  \section{Universality aspects and conclusions}
\label{sec:S5}
 In the final stage of this note, time is ripe now to glean the consequences for universality of our
  Pfaffian $\PVI$ characterization of the
  persistence probability. We shall also spice this up with a  conjecture, that we view as extremely plausible, corresponding to Eq.~(\ref{TB}), and that has probably already been proved in the conformal field theory or integrable literature by more knowledgeable people, 
  although 
  we have not managed to  track it down, or simply to understand the corresponding result had we read it.

 The first part of our considerations
  on universality is an immediate (and exact)
  consequence of our Fredholm Pfaffian generating function expression Eq.~(\ref{resP0I}), when rewritten in terms of the square-root of the determinant of a $2\times 2$ matrix block operator $\KK_2$ as in \cite{TWGOSE}. Copying  Eq.~(7) from \cite{TWGOSE}, replacing simply there the GOE sine-kernel by our sech-kernel, one has in coordinate-free, short-hand notations:
  \be
\label{defK2}
P_0^{\mathrm{HalfI}}(T;m) = \la \det{\lb \ID_2 -(1-m^2)\chi_T
\begin{pmatrix}
K \hfill & D K \hfill \\
 \varepsilon K - \varepsilon \hfill & K \hfill
\end{pmatrix}
\chi_T
\rb}
\ra^{1/2},
\ee
where $\varepsilon(x,y)=(1/2)\sgn(x-y)$, $D=\dd/\dd x$, $K$ is the kernel Eq.~(\ref{defK}), and $\chi_T$ the characteristic function of the interval $[-T/2,T/2]$.
The right-hand-side is  nothing but the Fredholm Pfaffian of $\mathbb{J}_2+(1-m^2)(J \KK_2)$, with  $J= \begin{pmatrix}
0 \hfill & 1 \hfill \\
 -1  \hfill & 0 \hfill
\end{pmatrix}$.
Expanding the gap probability generating function, the resulting $n$-point correlation functions
  $\rho_n(x_1,x_2,\dots,x_n)$ are found to coincide with the same Pfaffian
  expressions as those discovered in \cite{MS13}
  for the intensities of the Pfaffian Gaussian process $\{f(u)\}_{-1 < u < 1}$
  formed by the limiting random real Kac polynomial conditioned by $f(u)=0$. This holds up to the by-now standard changes of variable \cite{SM07,SM08,DM15} $u \approx 1-1/t \approx 1-e^{-T}$, that permits to identify in the vicinity of $u=1$ the zero-crossings properties of this limiting random Kac's polynomial, to those of
  the $2d$-randomly diffusing field $\{X(t)\}$ for $t\gg 1$, or equivalently those for large $T$ of the stationary process $\{Y(T)\}$ with correlator $A(T)=\sech{(T/2)}$. Resumming back
  the gap probability generating function specialized to $m=0$, we obtain from Eq.~(\ref{resP0I}) and then from Eq.~(\ref{F1decay}):
  \bq
 \label{threesixteenth}
P^{\mathrm{HalfI}}_0(T;m=0) = P^{\mathrm{2Diff}}_0(T | Y(0)=0)  \Longrightarrow \frac{\theta(0)}{2}=\tilta(2)=\frac{3}{16}.
\eq
  A subtle but important point concerns the  conditioning to an initial value $Y(0)=0$ appearing in Eq.~(\ref{threesixteenth}).
It can be read off from the equality of Eq.~(\ref{F1}) relating $F_1$ to its density $p_1$, and by a comparison
 with Gaudin's result 
for the  bulk GOE gap spacing distribution, $p^{\mathrm{GOE,bulk}}_1(\ell)=\frac{\dd^2}{\dd \ell^2} \det{(\ID - K_{\mathrm{sine},e}|_{[0,\ell]})}$, which involves instead the second-derivative, 
\textit{twice-conditioned} 
spacing distribution between eigenvalues \cite{Gaudin61, TWIntro}. In fact,  this once-conditioning of the first-passage probability $P_0(T)$ to the initial value $Y(0)=0$ was implicit in our approach: it corresponds, for an otherwise stationary process, invariant under $T \to -T$, in making a choice for the origin of the logarithmic timescale 
$T$ where the last-crossing of zero occurred, and to measure a (positive) length of time  from it. Notice that in this respect, 
Eq.~(\ref{ODEH1}) is left unchanged under 
$H(T) \to -H(-T)$. In perhaps more intuitive physical terms, the choice for the origin of $T$ and its sign ("the arrow of time")
is unambiguously set by the microscopic timescale $t_{mic} \sim 1$ beyond which the underlying
nonequilibrium coarsening process establishes, and that is thereafter rendered stationary by viewing it on the logarithmic timescale.

 Eq.~(\ref{threesixteenth}) therefore yields the proof of a long-standing conjecture \cite{Old} for the value of the persistence exponent of the $2d$-diffusing random field and of allied problems, a
result that had actually  been obtained before and independently
a few months ago while this work was in progress in the remarkable paper \cite{PS},
using  notably the connection with  truncated random
orthogonal matrices. The  techniques used by  Poplavskyi and Schehr also allow them 
to derive the equality as temporal processes between the sign 
of the  (conditioned) $2d$ diffusing field,  and the instantaneous value 
of the Ising spin located at the origin of a semi-infinite chain with symmetric initial
conditions, which entails in particular
the first equality of Eq.~(\ref{threesixteenth}).
 While we were aware since March 2018 that these authors had made this breakthrough, 
our work had followed a completely different path and,
as far as we can tell, none of the results presented here relies
on any  reported there. 
Notice in particular that in our approach the exact  $\tilta(2)=3/16$ 
is an ancillary consequence of the  tail behavior 
in the magnetization-symmetric case of our full Pfaffian $\PVI$ scaling distribution,
whose nature was not identified in \cite{PS}, let alone the geometric interpretation for $\theta(0)/2$, Eq.~(\ref{kappa}), unveiled.
Our expressions for the full scaling function valid at all times and for all $m$ should also grant access to other quantities than those studied in \cite{PS}. In this respect, note that our approach allows to recover immediately the algebraic decay for $s<0$ of $\langle e^{s{\cal{N}}_n} \rangle \propto 1/n^{\theta(e^s)}$, 
the exponential generating function for the number ${\cal{N}}_n$
of real zeroes
of  Kac's polynomials, that was computed explicitly with 
rather involved Pfaffian techniques in \cite{PS}, 
through 
the simple correspondence 
$n \mapsto \ln{(T/2)}$ \cite{SM07,SM08},  $e^s \mapsto m$ in Eq.(\ref{F1decay}).

At any rate, for a symmetric initial condition $m=0$, our distribution $F_1(T)$
represents  the
exact universal scaling first-passage probability distribution $P_0(T |Y(0)=0)$ for all Gaussian Pfaffian processes with a stationary sech-kernel correlator, explicitly given
by Eq.~(\ref{F1}), with $H(T)$ related to the $\PVI$ function $y(x)$, Eq.~(\ref{P6}), through Eq.~(\ref{Hy}).

The second part of our conclusions deal with the tentative exploitation of another remarkable
correspondence established recently in an impressive series of works \cite{Oleg1,Oleg2,Oleg3,Oleg4} (see also
\cite{BPZ84}, Eq.~(5.17), and \cite{Novikov}), and that
relates a generic $\PVI$  to conformal field theories with central charge $c=1$. Briefly, a tau-function for $\PVI$ provides the generating function of conformal blocks of four Virasoro primary fields, with
conformal dimensions $\Delta_\nu=\vartheta^2_\nu/4$ in terms of the monodromy parameters. Using our  Eq.~(\ref{monom})  gives here, up to any permutation of the four fixed singularities:
\be
\label{DT}
\la \Delta_1,\Delta_2,\Delta_3,\Delta_4\ra=\la 0,0,\frac{1}{16},\frac{1}{16}\ra, 
\ee
in which one is told to recognize the so-called sine-Gordon quantum field theory at the 
free-fermion point, itself  equivalent to the square of the $2d$  (static) Ising model at the critical temperature,
admittedly another famous Pfaffian process. More prosaically, we have noticed that the very same
sech-kernel, Eq.~(\ref{defK}), occurs systematically in  quantum field-theoretical approaches
for the $2d$ Ising model: it represents the so-called "heart of the kernel" in form-factor expansions, see e.g. \cite{BB92}. Contemplating some of the expressions in the latter work or in others \cite{BLC,WidomToda}, as well as the numerical values present in Eq.~(\ref{DT}) and Eq.~(\ref{P6}), and trying to figure out (literally) all this, we speculate that our $\PVI$ tau-function $F_2(T)$ corresponds,
 for the critical $2d$ Ising model  with appropriate boundary conditions (a certainly crucial point that we deliberately overlook), to an average between  two spins operators, $\bf{\sigma}$, and a \textit{disorder operator} $\bf{\mu}$ \cite{Kadanoff71}, i.e. a domain wall, each having at criticality conformal dimensions $1/16$. Recalling that the rational variable $x=e^{2T}$ for our $\PVI$ then
 also corresponds to the cross-ratio of the location of the four operators with dimensions
 $\Delta_\nu$, the above guess gives, for large separation $x$,
 \be
 \label{back}
 F_2(T)  \propto \langle 1 \sigma \sigma \mu \rangle_{{\mathrm{Ising}\, \mathrm{T}_c}} \propto \frac{1}{|x|^{3/16}} \propto e^{-3 T/8} \propto (t_1/t_2)^{3/8},
 \ee
a back-of-the-enveloppe argument that at least has the virtue of giving the correct answer for $\theta(0)$.
Believing in Eq.~(\ref{back}), and since in terms of the Onsager-Yang spontaneous magnetization critical exponent $\beta=1/8$, the conformal dimension at criticality
of both spin- and disorder-operator is $\beta/2$ , this would relate
in a noticeable fashion the  nonequilibrium persistence exponent  for  the $1d$-Ising model (with symmetric initial conditions) to ordinary static critical behavior for the $2d$-Ising model:
\be
\label{TB}
\theta(0) = 3 \beta=3/8.
\ee
Given the way the $m$-dependence appears in Eq.~(\ref{thetam}), it could even be possible to recover the general expression for $\theta(m)$  by twisting properly each of the two underlying $2d$-Ising models in the  conformal field theory description,
in order to accomodate the $m$-dependence through the corresponding 
continuously-varying magnetization exponent in some wedge-shaped geometry \cite{Cardy84}.
More generally, we believe that methods of boundary conformal field theory should be helpful in determining the critical properties of other unsolved nonequilibrium $1d$-reaction-diffusion processes, for instance the one studied in \cite{Monthus}, by unfolding in one more space dimension a  temporal quantity corresponding to a (non-local) observable here, and connecting it to a suitable and simpler observable up there.

As an overall general conclusion, 
the specific $\PVI$ probability distribution functions exhibited here 
in the context of persistence provide a particularly vivid example
in the emerging field of  \textit{stochastic integrability} \cite{Spohn}
or \textit{integrable probability} \cite{Borodin16}.
There are also other indications  \cite{Cras} that this is just the
tip of the iceberg of many other non-trivial universal limit distributions for
correlated random variables involving
Painlev\'e transcendents that remain to be discovered.
It is also hoped that their universal behavior 
 will one day incarnates
into genuine nonequilibrium physical systems, with experiments as striking as
those conducted  for KPZ-growing interfaces associated to the  $\PII$ Tracy-Widom distributions
\cite{Kaz10,Kaz11,Kaz12}. The correspondence between a generic $\PVI$ and Bonnet surfaces established in 
\cite{RC16,RC17}
should also
explain \cite{IDPrep2}, by the classical confluence along the Painlev\'e hierarchy all the way down to $\PII$, why geometry-dependent universality classes are observed in KPZ growth, where these nonequilibrium
interfaces
forever remembers the initial curvature they had.

\textit{Acknowledgements.} I warmly thank Clo\'e Colleras for making me believe this work was possible, as well as 
 my retired yet still active colleague and former neighboring office-mate 
Robert Conte for  sharing generously his  operative  knowledge of the old school Painlev\'e literature, and in particular for recognizing in Eq.~(\ref{ODEH1}) 
the mean curvature of a $\PVI$ Bonnet surface.
I also thank  Bertrand Delamotte for hosting
me at will in his lab, the LPTMC, in a stimulating environment,
and Gr\'egory Schehr for  keeping me informed of the advancement of his work.
Vincent Pasquier  gave me the opportunity to present an oral version of this work at IPhT, Saclay  on October 1st, 2018, where most of the results presented here were announced
\cite{Seminar}. I thank him for his keen interest, as well as
 comments from him and Didina Serban who helped me to figure out some of the tentative conformal implications. I also thank  Paul L. Krapivsky for reminding me 
 on this occasion of the existence of continuously-varying survival exponents for certain reaction-diffusion processes.
Last but not least, I acknowledge the support of  the CEA, my host institution, and of its local hierarchy at SPEC, for giving me the freedom over the past few years to pursue this endeavor.

\end{document}